\documentclass{elsart}

\usepackage{natbib}

 \usepackage{graphics}
\usepackage{amssymb}

\begin{document}

\begin{frontmatter}

% Title, authors and addresses

% use the thanksref command within \title, \author or \address for footnotes;
% use the corauthref command within \author for corresponding author footnotes;
% use the ead command for the email address,
% and the form \ead[url] for the home page:
% \title{Title\thanksref{label1}}
% \thanks[label1]{}
% \author{Name\corauthref{cor1}\thanksref{label2}}
% \ead{email address}
% \ead[url]{home page}
% \thanks[label2]{}
% \corauth[cor1]{}
% \address{Address\thanksref{label3}}
% \thanks[label3]{}

\title{The Effect of Use and Access on Citations}

% use optional labels to link authors explicitly to addresses:
% \author[label1,label2]{}
% \address[label1]{}
% \address[label2]{}

\author{Michael J. Kurtz},
\ead{kurtz@cfa.harvard.edu}
\ead[url]{www.cfa.harvard.edu/$\sim$kurtz}
\author{Guenther Eichhorn},
\author{Alberto Accomazzi},
\author{Carolyn Grant},
\author{Markus Demleitner},
\author{Edwin Henneken},
\author{Stephen S. Murray}
\address{Harvard-Smithsonian Center for Astrophysics, 60 Garden Street, 
Cambridge MA 01238, USA}

\begin{abstract}
% Text of abstract

It has been shown (S. Lawrence, 2001, Nature, 411, 521) that journal
articles which have been posted without charge on the internet are
more heavily cited than those which have not been.  Using data from
the NASA Astrophysics Data System (ads.harvard.edu) and from the ArXiv
e-print archive at Cornell University (arXiv.org) we examine the
causes of this effect.

\end{abstract}


\end{frontmatter}

% main text
\section{Introduction}
\label{intro}

During the past decade substantial changes have occurred in the manner
in which technical and scholarly literature is accessed and read.
While differing in detail from other disciplines, astronomy is broadly
typical of the developments, and can be used as an example.

Astronomers now have near total unimpeded direct electronic access to
nearly every important research article in astronomy beginning before
it is published, to the entire historical literature back to the
beginning of the nineteenth century, and to all the tabular numerical
data contained in the modern article. These developments have had many
sources, but four main developments stand out: The NASA Astrophysics
Data System (hereafter ADS) \citep{1993adass...2..132K,
2000A+AS..143...41K}, The ArXiv e-Print archive (hereafter ArXiv)
\citep{2001eps2.conf.....G}, the on-line journals \citep[beginning
with the Astrophysical Journal (Letters),][] {1995VA.....39....7D},
and the on-line data tables \citep[beginning with Astronomy \&
Astrophysics,][]{1995VA.....39..227O, 2000A+AS..143....1G}.  All these
new services (and several others) are interconnected by a complex
system of close collaborations and hyperlinks
\citep{1998lisa.conf..107B}.

The tremendous use of these new forms of access has been well
documented \citep[e.g.][]{2004JASIS........2K, 2004ArXiv}, and new
bibliometric measures which make use of them are beginning to be used
\citep[e.g.][]{2004JASIS.........K, 2003DLib....9....5B}.  The
question arises as to how, and whether, these new usage modalities
have effected the practice of citation, long the primary bibliometric
indicator of the usefulness of an academic article.

Recently \citet{2001Natur.411..521L} and \citet{2004npoa.conf.....B}
have demonstrated that articles which are available on-line at no
charge are cited at substantially higher rates than those which are
not. \citet{2004npoa.conf.....K} has shown that restrictive access
policies can cut article downloads to half the free access rate.

There are (at least) three possible, and non-exclusive, explanations
for the effect noted by \citet{2001Natur.411..521L} and
\citet{2004npoa.conf.....B}.  1.  Because the access to the articles
is unrestricted by any payment mechanism authors are able to read them
more easily, and thus they cite them more frequently; the Open Access
(OA) postulate.  2.  Because the article appears sooner it gains both
primacy and additional time in press, and is thus cited more; the
Early Access (EA) postulate.  3. Authors preferentially tend to
promote (in this case by posting to the internet) the most important,
and thus most citable, articles; the Self-selection Bias (SB)
postulate.

In this paper we present the results of two experiments designed to
distinguish to what extent each of the postulated explanations holds
true, for the astronomy literature.

\section{Data}
\label{data}

\subsection {The OA and EA Postulates}\label{dataoa}

Scholarly communication in astronomy is dominated by seven journals,
{\em The Astrophysical Journal, The Astrophysical Journal (Letters),
The Astrophysical Journal Supplement Series, The Monthly Notices of
the Royal Astronomical Society, Astronomy \&\ Astrophysics {\rm (which
merged with} Astronomy and Astrophysics Supplement Series {\rm in
2001)}, The Astronomical Journal, {\rm and} The Publications of the
Astronomical Society of the Pacific}.  Since the founding of the
pan-European journal {\em Astronomy and Astrophysics} thirty-five
years ago these journals have formed the core of the discipline.

In order to test the OA and EA postulates we use a data-set which
consists of all references from articles published in one of these
core journals to another article published in these journals,
beginning in 1970.  By restricting the citation universe to the core
journals we minimize the effects of systematics, such as changes in
the relative popularity of publishing in journals versus conference
proceedings, on the usage trends we measure.

These citation data are maintained in the ADS
database, and are nearly 100\%\ complete.  While the original source
for the ADS citation database was ISI, for the core astronomy journals
this has now been totally superseded by parsing the full-text of the
articles, as sent by the journals, or, for the older ones, parsing the
output of images of the article pages as analyzed by a character
recognition program \citep{1999adass...8..291A, 2004cs........1028D}.

\subsection {The SB Postulate}

To test the SB postulate (vs. the combined OA and EA postulates) we
use data concerning the 2592 articles published in 2003 by {\em The
Astrophysical Journal} contained within the ADS database.  We use the
per article citation information, and the concordance between {\em
Astrophysical Journal} articles and pre-prints in the ArXiv.

The choice of the data sample is somewhat arbitrary.  The sample must
sufficiently homogeneous that we do not measure differences in the
citation practices of different disciplines; it must be old enough to
have built up statistically significant differences between the more
and less cited articles; and it must be new enough to measure the
current behavior of ArXiv users.  This sample meets these criteria.

The citations are as of 15 August 2004 and are taken from all sources
known to the ADS. We use the entire data-set to maximize the signal to
noise; because the cited articles are quite homogeneous, and the
citation period is relatively short, there are no systematics which
would require us to restrict the data set of citing articles.

The concordance is created by: 1. Parsing the
information submitted by the authors to the ArXiv (yielding 70\% of
all 2003 {\em ApJ} articles); and 2. Comparing the title and author
information for the ArXiv publication with the {\em Astrophysical
Journal} articles author and title information, using any additional
author supplied information (yielding 4\% of all 2003 {\em
ApJ} articles).

As the ArXiv preprint can (and often does) have a different title than
the corresponding journal article, indeed it can have a different
author list, it is not possible to have a 100\%\ complete concordance.
By making spot checks, and by looking at each of the 200 most cited
papers in detail, we estimate that we could be underestimating the
fraction of ArXiv articles which correspond with {\em Astrophysical
Journal} articles by 2\%, but none of these are in the top 200 cited
articles.  Finding these (if present) would not significantly change
any of our conclusions.

\section{Technique}
\label{technique}

\subsection {The OA and EA Postulates}

To test the OA and EA postulates we make note of the fact that
substantial changes in ability of astronomers to access research
articles occurred during the decade of the 1990s.  During this period
the ArXiv was founded \citep{2001eps2.conf.....G}, it has since grown
to where the majority of research articles in astronomy are submitted
to it.  Also during this period the ADS was founded
\citep{1993adass...2..132K} and during the 36 month period ending in
1999 it (with the kind permission of the journals) scanned and put
on-line all the back issues of the seven (then eight) core astronomy
journals.  In both cases the articles are available without cost
through an easy to use web interface.  Both services are heavily used
\citep{2004JASIS........2K, 2004ArXiv}.

If increased access effects the citation rate we should be able to see
changes in the citation rates for older articles and for very new
articles beginning in the mid to late 1990s.  For the older articles
this tests the OA postulate, as these articles have long since
appeared, but are suddenly now available without charge, and without
the need to go to a library.  For the very new articles this tests a
combination of the OA and EA postulates.  The articles are available
without charge, but also substantially before they are available in
the journals.

We create a statistic to measure the citation rate which is
insensitive to changes in the number of articles published during
different periods, the probability $P(t,t_0,\Delta t)$ that an article
published at time $t$ will cite a particular article published during
the period between $t_0$ and $t_0 + \Delta t$ ago; thus $P(t,10,10)$
is the probability that an article published at time $t$ will cite an
article published between 10 and 20 years before $t$.

$$
P\left(t,t_0,\Delta t\right) \equiv \frac{citations(t,t_0,\Delta t)}{N(t)*N(t,t_0, \Delta t)} 
$$
{\rm where} \hfill \nonumber

$citations(t,t_0,\Delta t)$ {\rm is the number of citations in papers
at time} $t$ {\rm to papers published between} $t_0$ {\rm and} $t_0 +
\Delta t$ {\rm before time} $t$. \nonumber

$N(t)$ {\rm is the number of papers published at time} $t$.  \nonumber 

$N(t,t_0,\Delta t)$ {\rm is the number of papers published in the
period between} $t_0$ {\rm and} $t_0+\Delta t$ {\rm before time}
$t$. \nonumber

We use a one month granularity throughout, so $N(t)$ is the number of
articles published in a one month period of time, etc..

\subsection {The SB Postulate}

To test the SB postulate we note that the magnitude of the combined OA
and EA effects makes a clear prediction concerning the relative
distribution of articles which had been submitted to the ArXiv versus
articles which had not in a list of recent papers sorted by number of
citations.  The higher the ratio of citation rates between
ArXiv-submitted articles and non-ArXiv-submitted articles the lower
the fraction of non-ArXiv-submitted articles expected in the top 100
or 200 cited articles in the sample.

Given an ArXiv/non-ArXiv citation ratio, and the unbiased frequency
distribution of citations one can easily determine the probability
that a particular number of non-ArXiv submitted papers will be in the
top 100 or 200 most cited papers by Monte Carlo simulation.  Comparing
these probabilities with the actual measured number allows one to
accept or reject the hypothesis that OA+EA effects alone can account
for the measured distribution of highly cited articles.

To estimate what the relative distribution of citations for
ArXiv-submitted and non-ArXiv-submitted articles would be were there
no SB effect we must estimate what the total citation distribution
would be in the absence of any SB effect.  While we could use a Lotke
type power law we choose to estimate this, and all other elements in
our stochastic model, based very closely on the actual data for the
2003 {\em Astrophysical Journal}.

For the ArXiv/non-ArXiv citation ratio we take the actual measured
value; for the unbiased citation frequency distribution we take the
actual citation frequency distribution.  Next we randomly choose N
papers, where N is the number (683) of non-ArXiv submitted papers in the
sample, and we reduce the citation counts for these papers by the
measured citation ratio. (We could also have increased the citation
counts for the 1909 articles which represent the ArXiv submitted
set, because we are only interested in the relative rank these
procedures are equivalent).

Finally we
sort the modified counts and determine how many of the modeled
non-ArXiv submitted papers are in the top 100 and 200 most cited
papers.  We repeat this procedure many times, to acquire the
statistical distribution.

Finally we compare the actual measured number of non-ArXiv submitted
papers in the top 100 and 200 most cited papers with this expected
distribution.

\section{Results}
\label{results}

\subsection {The OA and EA Postulates}

Figures 1 and 2 show the results of calculating the $P$ statistic for
a number of different offset times, using the citation data described
in section \ref{dataoa}..  Figure 1 shows the older material, with
$P(t,10,10)$ at the bottom, followed by $P(t,5,5)$, the $P(t,3,2)$,
and finally $P(t,2,1)$ at the top. The decrease in the value of the
$P$ with age is explained by the obsolescence function
\citep[e.g.][for the astronomy literature]{2004JASIS.........K}.

\begin{figure}
\vspace{60mm}
\centering
\includegraphics{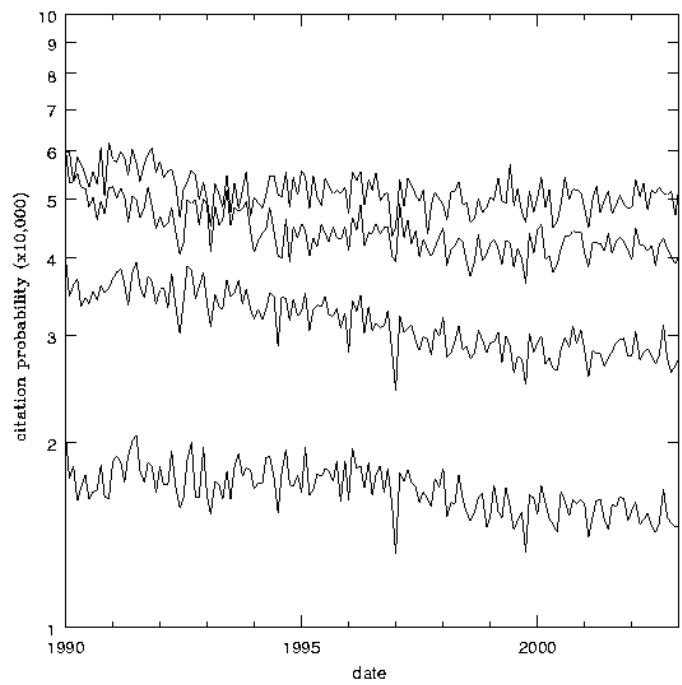}
\caption{The $P$ statistic for older material.  From the bottom:
$P(t,10,10), P(t,5,5), P(t,3,2), P(t,2,1)$.}
\end{figure}

Figure 2 shows the $P$ statistic for more recent material, with
$P(t,0,0.5)$ on the bottom, followed by $P(t,0.5,0.5)$, then $P(t,1,1)$
on top.  Here the increase in $P$ with age is fully consistent with
the rapid increase in citation rate immediately following publication
\citep[e.g.][for the astronomy literature]{2004JASIS.........K}.

\begin{figure}
\vspace{60mm}
\includegraphics{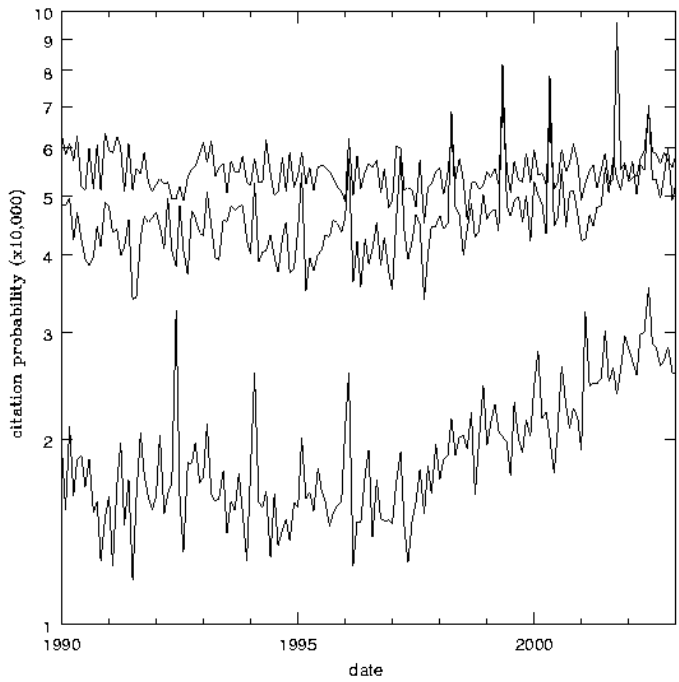}
\caption{The $P$ statistic for newer material.  From the bottom:
$P(t,0,0.5), P(t,0.5,0.5), P(t,1,1)$.}
\end{figure}

The most striking feature visible on these plots, the large spikes in
the short time lag data shown in figure 2, are not important to this
discussion.  They are caused by the regular publication of Trimble and
Ashwanden's (e.g. 2004 and many previous) review of the literature in
astrophysics published during the preceding calendar year.

Looking first at figure 1 we see that the various $P$ functions are
essentially flat over the 13 year period shown in the plots, trending
slightly lower for the older papers.  There is no feature of any kind
which can be associated with the change in article access status
caused by placing on-line, and without charge scanned versions of the
full-text during the late 1990s.

Figure 2 shows a different story.  Here both $P(t,0,0.5)$ and
$P(t,0.5,0.5)$ show a substantial change in slope beginning in the mid
1990s and corresponding with the advent and gradual increase in
popularity of the ArXiv service.  With the exception of the Trimble
and Ashwanden spikes $P(t,1,1)$ is essentially flat.

\subsection {The SB Postulate}

Figure 3 shows the distribution of the number of non-ArXiv submitted
articles in the top 200 most cited articles using the actual data for
the 2003 {\em Astrophysical Journal} articles to build the model.  The
measured ArXiv/non-ArXiv citation ratio for these data is 2.11, there
are 1909 ArXiv submitted articles and 683 non-ArXiv submitted
articles.  The peak of the distribution is at 16, 16 is the number of
non-ArXiv submitted articles in the top 200 most cited papers with the
highest expectation value under the assumption that there is no SB effect.

\begin{figure}
\vspace{60mm}
\includegraphics{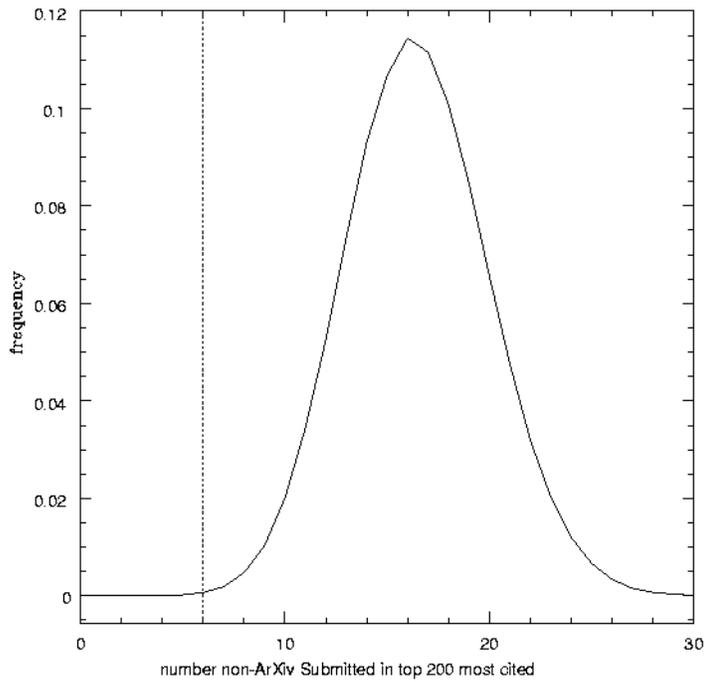}
\caption{The probability distribution for the number of non-ArXiv
submitted papers in the top 200 most cited papers, where there are
1909 ArXiv submitted papers, 683 non-ArXiv submitted papers, and the
ArXiv/non-ArXiv citation ratio is 2.11, under the assumption that 
there is no SB effect}
\end{figure}

The dotted line, at 6, shows the actual number of non-ArXiv submitted
articles in the top 200 most cited 2003 {\em Astrophysical Journal}
articles.  Six or fewer articles appear in the model distribution
fewer than one time in 1000; we can thus reject the hypothesis that
the citation distribution of ArXiv submitted vs. non-ArXiv submitted
articles is caused by the ArXiv/non-ArXiv citation ratio (i.e. by the
combined EA+OA effect) alone at the 99.9\%\ confidence level.  A
similar analysis of the top 100 most cited papers, where there is only
one non-ArXiv submitted paper, rejects this hypothesis at the 99.8\%\
confidence level.

We thus must accept that there is some other systematic effect,
correlated with citation count, which affects the ArXiv submitted
vs. non-ArXiv submitted status of a paper.  This is (without implying
anything about motivations) the SB postulate.

\section{Analysis}
\label{analysis}

Figure 1 shows clearly that increasing access to existing articles
does not increase the probability that they will be cited.  This
implies that there is no significant population of astronomers who are
both authors of major journal articles and who do not have
``sufficient'' access to the core research literature.  This also
implies that increasing access above a ``sufficient'' level has no
influence on citation frequency.  Because the number of downloads
exceeds the number of traditional paper reads by a factor of several
\citep{2004JASIS.........K} this suggests that there are differences in how
researchers use electronic articles compared with how paper articles
were used in the past.  What these differences are is beyond the scope
of this article.

Figure 2 shows to what extent the ArXiv has effected the citation of
very recent articles.  Notice that the value of $P(t,0,0.5)$ at the
end of 2003 is much closer to the value of $P(t,0.5,0.5)$ at the
beginning of 1990, and that $P(t,0.5,0.5)$ at the end of 2003 meets
$P(t,1,1)$, which is constant throughout the 13 year period.  This can
be understood as changing the mean effective publication date by about
five or six months.  This would imply that the peak in the article
age---citation rate diagram would be shifted toward younger articles,
an effect already seen by \citet{2003sinn....3.....B}.

The results of the simulations shown in figure 3 indicate (at the
99.9\%\ confidence level) that there must be some sort of selection
based on article quality in determining which papers are submitted to
the ArXiv, and which not.  Because papers in the ArXiv are not
refereed (except for a kook filter) this suggests that authors
self-censor or self-promote, or that for some reason the most citable
authors are also those who first use the new publication venue.

Taken together these figures suggest that, in astronomy, there is a
strong EA effect and a strong SB effect; there is also no indication
of any OA effect.  At first this seems counterintuitive; if more
people could read a document one might expect that more people would
then cite it.

We suggest that the basic reason why there seems no OA effect in
astronomy is that for a person to be in the position to write an
article for a core astronomy journal that person must already be in a
position to read those journals, and must also be in a position to
perform astronomical research.  In terms of barriers to entry into the
astronomical research community the second requirement is much larger
than the first.  Because the marginal cost of being an astronomer with
access to the core literature is so much lower than the cost of being
an astronomer in the first place, it is reasonable to postulate that
essentially all astronomers have access to the core literature through
existing channels, and thus do not require an OA alternative path in
order to read and cite articles.  

We believe that the claims that the citation rate ratio of papers
openly available on the internet (via ArXiv or some other mechanism)
vs those not available through those means is caused by the increased
readership of the open articles \citep[][this is sometimes called the
Lawrence Effect, or the OA advantage]{2001Natur.411..521L,
2004npoa.conf.....B} are somewhat overstated, especially for well
funded disciplines with high barriers to entry.  The ongoing debate
concerning the desirability of Open Access literature
\citep[e.g.][]{2004Natur.onlinedebate, 2004Amsci.forum} ought not be
affected by this result, as once all literature is open access, there
can no longer be a differentiation between open and non-open articles
anyway.

Before open access to the core literature in astronomy could have a
large effect on citation practices there would also need to be open
access to the basic data of astronomical research, so as to lower the
entry barrier to be close to just time and ability.  There is a
substantial effort in astronomy currently to do just that
\citep[e.g.][]{2001vof..conf.....B,2004tivo.conf.....Q}.

\section{Acknowledgments}

MJK wishes to thank S. Harnad and T. Brody for substantial e-mail
discussions; and wishes to thank P. Ginsparg for discussions and
hospitality.  We also than an anonymous referee for adding to the
clarity of the presentation.

The ADS is supported by NASA under grant NCC5-189.

% The Appendices part is started with the command \appendix;
% appendix sections are then done as normal sections
% \appendix

% \section{}
% \label{}

% Bibliographic references with the natbib package:
% Parenthetical: \citep{Bai92} produces (Bailyn 1992).
% Textual: \citet{Bai95} produces Bailyn et al. (1995).
% An affix and part of a reference:
%   \citep[e.g.][Ch. 2]{Bar76}
%   produces (e.g. Barnes et al. 1976, Ch. 2).

\end{document}